# Synthesis of Group-IV ternary and binary semiconductors using epitaxy of GeH$_3$Cl and SnH$_4$

Aixin Zhang, [1] Dhruve A. Ringuala,[1] Matthew A. Mircovich,[2] Manuel A. Roldan,[3] John Kouvetakis[1], and José Menéndez[2]

[1]*Arizona State University, School of Molecular Sciences, Tempe, AZ, USA 85287-1604*

[2]*Arizona State University, Department of Physics, Tempe, AZ, USA 85287-1504*

[3]*Eyring Materials Center, Arizona State University, Tempe, AZ 85287 USA*

**ABSTRACT**

Ge$_{1-x-y}$Si$_x$Sn$_y$ alloys were grown on Ge buffer layers at ultra-low temperature using reactions of SnH$_4$ and GeH$_3$Cl for the first time. The latter is a newly introduced CVD source designed for epitaxial development of group IV semiconductors under low thermal budgets and CMOS compatible conditions. The Ge$_{1-x-y}$Si$_x$Sn$_y$ films were produced between 160-200°C with 3-5% Si and ~ 5-11 % Sn, which traverses the indirect to direct gap transition in Ge-Sn materials. The films were fully strained to Ge and exhibited defect-free microstructures, flat surfaces, homogeneous compositions, uniform thicknesses and sharp interfaces as required for device manufacturing. A comparative study was then conducted to investigate the applicability of GeH$_3$Cl for the synthesis of Ge$_{1-y}$Sn$_y$ binaries under similar experimental conditions. The Ge$_{1-y}$Sn$_y$ films were grown fully strained to Ge, but with reduced Sn compositions ranging from ~ 2 - 7 % and lower thicknesses relative to Ge$_{1-x-y}$Si$_x$Sn$_y$. This prompted efforts to further investigate the growth behavior of Ge$_{1-y}$Sn$_y$ using the GeH$_3$Cl method, bypassing the Ge buffer to produce samples directly on Si, with the aim of exploring how to manage interface strain. In this case the Ge$_{1-y}$Sn$_y$ on Si films exhibited compositions and thicknesses comparable to Ge$_{1-y}$Sn$_y$-on-Ge films; however, their strain states were mostly relaxed, presumably due to the large misfit between Ge$_{1-y}$Sn$_y$ and Si. Efforts to increase the concentration and thickness of these samples resulted in non-homogeneous multi-phase materials containing large amounts of interstitial Sn impurities. These outcomes suggest that the incorporation of the small Si atoms in the tetrahedral structure of Ge$_{1-x-y}$Si$_x$Sn$_y$ might compensate for the large Ge-Sn mismatch by lowering bond strains. Such an effect reduces strain energy relative to the binaries, enhances structural stability, promotes higher Sn incorporation and increases the critical thickness that can be achieved. The work described in the paper demonstrates for the first time the use of a novel GeH$_3$Cl precursor to fabricate Ge$_{1-x-y}$Si$_x$Sn$_y$ semiconductors at ultra-low temperatures akin to MBE processes. Furthermore, it illustrates that addition of Si into the lattice of Ge$_{1-x-y}$Si$_x$Sn$_y$ leads to enhanced crystal growth and improved structural and morphological properties.

# INTRUDUCTION

GeH$_4$ (germane) is the commonly used reagent for fabricating Ge-containing semiconductors, including Ge-on-Si platforms for photonic applications as well as Ge$_{1-x}$Si$_x$ materials and devices on Si.[1,2] This compound represents the industry standard for manufacturing state-of-the-art technologies at a commercial scale. Most recently, GeH$_4$ has been used in combination with SnCl$_4$ for CVD of Ge$_{1-y}$Sn$_y$ and Ge$_{1-x-y}$Si$_x$Sn$_y$ alloys.[3-6] These new compounds extend the optoelectronic capabilities of the group-IV semiconductor family beyond the 1.5 µm Ge-threshold and well into the infrared, while maintaining compatibility with Si platforms.[7,8] Initial demonstrations of Si-Ge-Sn photodetectors with extended infrared responsivity[9] and lasers operating at low temperature[10] established the potential of these alloys for practical applications in the early days, even though these materials are thermodynamically stable only at very low Sn concentration of ~1%. As a result, their fabrication required non-equilibrium synthetic routes at low temperatures.[11-14]

The original CVD approaches to Si-Ge-Sn utilized higher-order Ge hydrides (Ge$_2$H$_6$, Ge$_3$H$_8$ and Ge$_4$H$_{10}$) instead of the simpler GeH$_4$.[15-19] This choice was made due to their higher reactivity at low temperatures. The latter are crucial for synthesizing materials with ultrahigh Sn contents, which are essential for obtaining continuous bandgap tuning across the entire SWIR and MWIR spectral ranges.[20,21] Ge$_2$H$_6$ was employed in the first CVD synthesis of Ge$_{1-y}$Sn$_y$ via reactions with SnH$_4$ and also in the initial demonstration of Ge$_{1-x-y}$Si$_x$Sn$_y$ alloys.[15,16] Subsequently Ge$_2$H$_6$ was applied to grow Ge$_{1-y}$Sn$_y$ using SnCl$_4$ as the source of Sn.[17] The motivation for developing Ge$_3$H$_8$ and Ge$_4$H$_{10}$ was the need to deploy lower temperature alternatives to Ge$_2$H$_6$. The higher-order polygermanes were expected to provide significant advantages over conventional CVD routes by displaying enhanced, tunable reactivities that can be optimized to deliver metastable materials conditions.[18,19,22-24] The advances enabled by Ge$_3$H$_8$ and Ge$_4$H$_{10}$ include: (a) extended compositional ranges reached by lowering the growth temperature to levels not achievable with other CVD approaches, allowing the growth of next-generation Ge$_{1-y}$Sn$_y$ and Ge$_{1-x-y}$Si$_x$Sn$_y$ samples with ultrahigh Sn contents up to 36% Sn, (b) compatibility with *in situ* doping with full activation of dopants and extremely flat doping profiles at low temperatures, and (c) nearly full incorporation of all group-IV elements in the gas phase into the solid film, minimizing waste of rare and expensive Ge. The only disadvantage of the Ge$_3$H$_8$ and Ge$_4$H$_{10}$ route is that their synthesis requires pyrolysis of Ge$_2$H$_6$ in a flow reactor at 250°C. This process involves extra steps that not only add to the cost of production but also generate significant amounts of GeH$_4$ secondary products. Fortunately, GeH$_4$ can be fully recovered and recycled, ensuring no net loss of Ge.

Due to its compatibility with commercial reactors, the GeH$_4$ approach to Sn-containing alloys remains particularly attractive for mass scale production compared to the above higher order Ge hydrides. However, despite early successes and enormous potential, the path for widespread deployment does not appear to be straightforward. One issue is the reactivity mismatch between GeH$_4$ and SnCl$_4$ under temperature conditions compatible with substitutional Sn-incorporation into Ge. Overcoming this mismatch requires a 100-1000 fold excess partial pressure of GeH$_4$ to produce meaningful growth of crystalline films under the metastable conditions employed.[3,6] This approach is wasteful and potentially unsustainable given the natural scarcity of elemental Ge and the rapidly increasing cost of commercial Ge hydrides.[25,26] Furthermore, the compositions in this regime are controlled mainly by the deposition temperature rather than the more deliberate use of classical stoichiometry of the gaseous sources. This reliance on temperature control makes mass production problematic due to reproducibility and reliability concerns.

We recently investigated alternatives to polygermanes (Ge$_3$H$_8$ and Ge$_4$H$_{10}$) for CVD of Si-Ge-Sn semiconductors targeting precursors that are more easily accessible in large quantities. The goal was to find compounds with physical properties closer to GeH$_4$ but with improved chemical reactivity for low temperature CVD. To this end we produced and tested the GeH$_3$Cl derivative of GeH$_4$.[27] The synthesis is conducted via straightforward chlorination of readily available GeH$_4$ at room temperature, based on the reaction

$$\text{SnCl}_4 + \text{GeH}_4 \rightarrow \text{GeH}_3\text{Cl} + \text{SnCl}_2 + \text{HCl} \qquad (1)$$

The compound was found to exhibit higher reactivity than GeH$_4$, essentially rivaling that of Ge$_4$H$_{10}$ which represents the most reactive chemical source from the general family of Ge hydrides. The potential of GeH$_3$Cl as a CVD sources was demonstrated in prior work with the deposition of device quality Ge films on Si wafers at 330-350 °C.[27] These films exhibited crystallinity, thickness and morphology comparable to films produced by Ge$_4$H$_{10}$. The layers were doped with donor atoms, allowing fabrication of Ge photodiodes exhibiting higher responsivity and lower dark current compared to Ge$_4$H$_{10}$ analogs. In addition, it was shown that GeH$_3$Cl is amenable to ultra-low temperature (200 °C) deposition of Ge$_{1-y}$Sn$_y$ films (2%Sn) on Si. These results suggested that the GeH$_3$Cl likely follows a similar dissociation pathway as Ge$_4$H$_{10}$, involving the formation of GeH$_2$ reactive intermediates. This was further borne out by control thermolysis experiments of GeH$_3$Cl. These showed that the compound undergoes homogeneous dissociation driven by HCl elimination, likely yielding GeH$_2$ according to

$$\text{GeH}_3\text{Cl} \rightarrow \text{GeH}_2 + \text{HCl} \qquad (2)$$

GeH$_2$ can then serve as the enabling building blocks for Ge crystal assembly at low temperatures. The high reactivity of GeH$_3$Cl motivated us to further develop this compound as a viable precursor for ultralow temperature synthesis of epitaxial films. In this study, we built upon our prior work on depositing pure Ge by using GeH$_3$Cl to synthesizing Ge$_{1-x-y}$Si$_x$Sn$_y$ alloys, demonstrating the versatility of this precursor in depositing a range of Ge-based semiconductors. The Ge$_{1-x-y}$Si$_x$Sn$_y$ samples are produced via reactions of GeH$_3$Cl with the SnH$_4$ and Si$_4$H$_{10}$ sources using Ge buffered Si platforms as substrates. The Sn contents was varied from 4-10 % and the Si from 5-2.5 % by reducing the temperature from 200 °C to 160 °C. Growth of Ge$_{1-y}$Sn$_y$ analogs was also performed under similar conditions to investigate the influence of Si incorporation on the growth properties of the binary. In this connection we find that Si atoms in the structure facilitate higher Sn substitution. This, in turn, promotes larger thicknesses in epitaxial layers and induces better crystallinity as revealed by thorough characterizations of the samples by XRD, TEM and RBS. Our collective results indicate that adopting GeH$_3$Cl in place of Ge$_4$H$_{10}$ as a Ge source for low temperature processing is both feasible and practical.

Notably, GeH$_3$Cl exhibits similar reactivity to SnH$_4$ as required for fabricating Sn containing samples with tunable compositions across a wide range. Furthermore, GeH$_3$Cl is a more versatile CVD precursor compared to Ge$_4$H$_{10}$ due to its much higher vapor pressure of 530 Torr at room temperature, whereas Ge$_4$H$_{10}$ has a vapor pressure of only 1 Torr at 22 °C. This drawback limits the practicality of Ge$_4$H$_{10}$ for large-scale industrial applications. A key benefit of GeH$_3$Cl is that, like Ge$_4$H$_{10}$, it reacts at ultralow temperatures that are compatible with the stability range of high Sn content alloys. However, GeH$_3$Cl offers additional benefits in that it is easier to handle and less expensive to produce compared to Ge$_4$H$_{10}$.

# EXXPERIMENTAL

## Growth of Ge$_{1-x-y}$Si$_x$Sn$_y$ via GeH$_3$Cl

The depositions of films were performed in a gas-source molecular epitaxy chamber (GSME) with base pressure $P$ of $10^{-10}$ Torr. The chamber typically operates at $T$ <400 °C and $P$ ranging from ~ $10^{-4}$-$10^{-7}$ Torr. Four-inch Ge-buffered Si wafers with 0.01 Ω·cm resistivity served as substrates. The buffer layers were grown at 350°C using GeH$_3$Cl or Ge$_4$H$_{10}$. Their thicknesses ranged from 500 nm to-700 nm.

The Ge-buffered substrates were chemically cleaned in a 5% HF/H$_2$O bath, dried under nitrogen, and loaded into the chamber. They were then heated to 650°C under UHV for several hours to desorb surface impurities. A glass vessel with liquid GeH$_3$Cl was connected to one gas inlet of the injection manifold of the chamber, while another vessel containing a 10:1 mixture of SnH$_4$ and Si$_4$H$_{10}$ was connected to a second inlet allowing independent control of the reactants inside the chamber. This dual arrangement ensured that the gaseous sources combined in desired proportions at the growth surface in a steady and continuous manner. GeH$_3$Cl vapor was admitted first inside the reactor, raising the pressure from $10^{-10}$ Torr to 4×$10^{-5}$ Torr. Then the SnH$_4$ and Si$_4$H$_{10}$ stock mixture was injected, further raising the pressure to a final 6×$10^{-5}$ Torr. The latter was held constant during the experiments by continuous turbo pumping of the chamber contents.

A series of growth experiments was conducted under these conditions by adjusting the reaction temperature from 200°C to 160 °C. While temperature was the primary variable, flux rates and reactant ratios were also adjusted to optimize crystal quality and Si/Sn stoichiometry. This approach provided insights into the reactivity behavior of GeH$_3$Cl with the Si and Sn sources, as well as the incorporation of Si and Sn atoms under ultra-low temperatures.

| sample | Tg(°C) | XRD Sn (%) | RBS Sn (%) | Si (%) | ε (%) | a$_0$ (Å) | Thickness (nm) |
|---|---|---|---|---|---|---|---|
| 1 | 195 | 5.5 | 5.4 | 5 | -0.4563 | 5.6919 | 70 |
| 2 | 183 | 6.7 | 6.3 | 4.6 | -0.6051 | 5.7024 | 100 |
| 3 | 180 | 7 | 6 | 4.1 | -0.7038 | 5.706 | 75 |
| 4 | 180 | 6.9 | 6.3 | 3 | -0.7119 | 5.7074 | 40 |
| 5 | 175 | 8.2 | 8.3 | 2.3 | -0.9283 | 5.72 | 57 |
| 6 | 170 | 8.2 | 7.8 | 2.5 | -0.9801 | 5.7215 | 55 |
| 7 | 165 | 9.6 | 9 | 2.5 | -1.147 | 5.7319 | 34 |
| 8 | 160 | 10.7 | 10.7 | 2.5 | -1.1888 | 5.7407 | 35 |

**TABLE 1.** Summary of Ge$_{1-x-y}$Si$_x$Sn$_y$ ($y$=0.055-0.11) samples grown on Ge-buffered Si(100) and corresponding growth temperature $Tg$, compositions, in-plane strain ($\varepsilon$), relaxed lattice constant ($a_0$) and thicknesses.

Samples with mirror like surfaces were produced under optimized conditions and were fully characterized by Rutherford backscattering (RBS), high-resolution x-ray diffraction (HRXRD), Atomic Force Microscopy (AFM), Cross-Sectional Transmission Electron Microscopy (XTEM) and Raman scattering. Table 1 summarizes growth conditions and characterization results for representative samples including, Sn/Si compositions, cubic lattice constants, strain states and film thicknesses. The bulk atomic concentrations of Si, Sn and Ge were directly measured by RBS and corroborated by XRD. RBS was essential for determining Si concentration which XRD alone cannot provide.

Figure 1(a) shows 2 MeV RBS spectra from sample 2 in Table 1. The logarithmic scale plot provides an enlarged view of the baseline, highlighting a clearly resolved Si signal indicated by the arrow. Strong Ge and Sn peaks from the buffer and epilayer are visible in the spectrum. The Si signal appears as a low intensity peak adjacent to the Ge signal from the buffer layer. Model fits indicate Si and Sn contents of 4.6% Si and 6.3% Sn, respectively, and a film thickness of 100 nm, in agreement with ellipsometry measurements. In addition to the 2 MeV spectra, higher energy plots at 3-3.7 MeV were also measured to resolve Si peaks in cases where the samples exhibited overlapping Ge and Si signals from the buffer and epilayer, respectively. We note that all samples were analyzed using RBS to verify the presence of distinct Si peaks, and model fits of the spectra provided the absolute Si contents listed in Table 1.

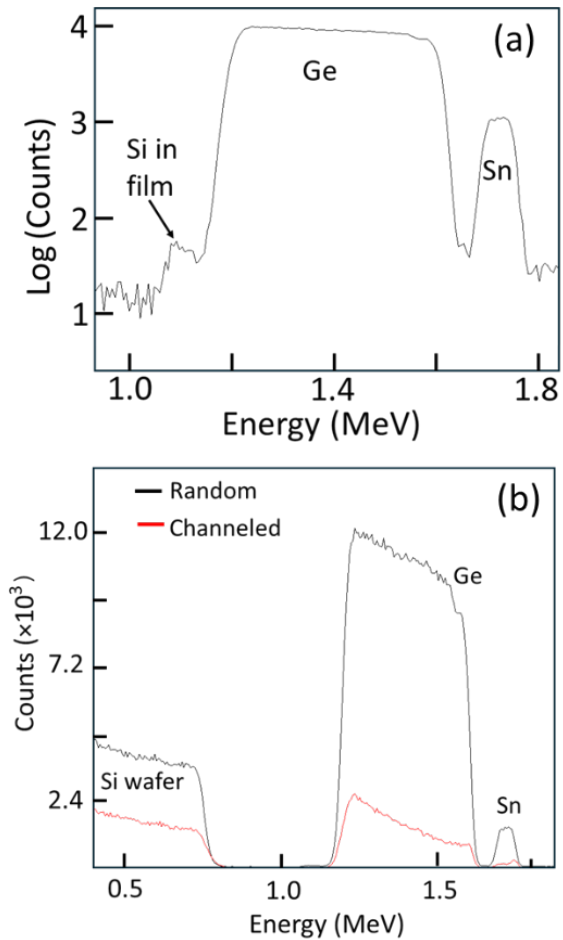

**Figure 1** (a): 2.0 MeV RBS spectrum of $Ge_{0.891}Sn_{0.063}Si_{0.046}$ plotted in a logarithmic scale to enlarge the Si signal. Panel (b): Random RBS (black line) and channeled (red line) spectra of same sample. The high degree of channeling indicates good crystallinity and complete Sn substitutionality in Si-Ge lattice.

In addition to the random RBS the channeled spectra were also measured. Figure 1(b) compares the random (black line) and channeled spectra (red line) of sample 2 measured at 2 MeV. The random spectrum features a broad peak resulting from overlapping Ge signals of the epilayer and Ge buffer, along with a narrow Sn peak from the film. The high degree of channeling indicated by the low intensity of the red signal suggests that the constituent atoms occupy the same lattice. This is consistent with the single-phase character indicated by the XRD analysis discussed below. The classic channeling profile observed throughout this sample is characteristic of a material with excellent crystallinity, well aligned interfaces and low defectivity microstructure.

The Sn content in the $Ge_{1-x-y}Si_xSn_y$ samples in Table 1 is increased from 5.5 % to 11% as the growth temperature is lowered from 200 to 160 °C, respectively. This trend aligns with the expected compositional dependence behavior in this class of materials, namely progressively lower temperatures required for full Sn substitution lead to higher Sn contents in the lattice. Additionally, the table indicates that decreasing the growth temperature also reduces the growth rate, which limits the overall thickness of the resultant layers as shown in the last column. This reduction in thickness may also be attributed to a concomitant increase of the strain differential between Ge and $Ge_{1-}$ yielded 4.93 % Si, 5.44% Sn, and 89.63 % Ge, while those of the green plots yielded 5.12 % Si, 5.52 % Sn, and 89.36 % Ge. These results indicate that the atomic concentrations are uniform with depth and closely reflect the $Ge_{0.895}Sn_{0.055}Si_{0.05}$ stoichiometry measured by RBS.

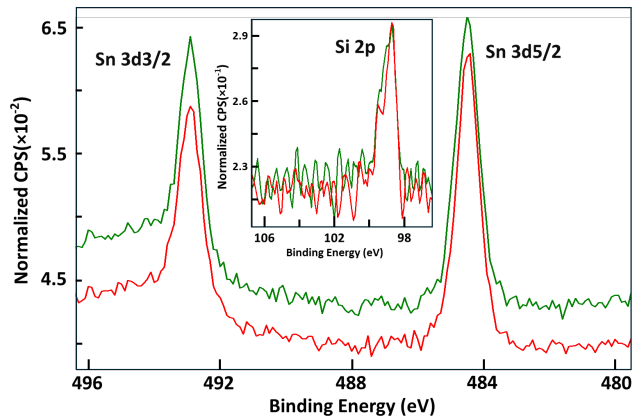

**Figure 2** The green and red plots represent two XPS spectra collected from the same $Ge_{0.895}Sn_{0.055}Si_{0.05}$ film. These were obtained after consecutive 25-second sputtering intervals, with green plot showing spectrum after first interval and the red after second interval. The main panel compares the Sn signals from the spectra. The peak shape and intensity are nearly identical but they are slightly offset for differentiation. Inset shows a near prefect overleap of the Si signals indicating closely matched compositions. Quantifications of the Sn and Si signals were performed to obtain stoichiometry of the sample using instrument specific sensitivity factors.

High resolution XRD was employed to characterize the epitaxial alignment and measure lattice constants of the films. Figure 3 shows on-axis XRD plots with strong, sharp (004) Bragg reflections for the epilayer and buffer for sample 1. The interference fringes on both sides of the $Ge_{1-x-y}Si_xSn_y$ (004) peak indicate a sharp interface with Ge. The inset of the figure presents (224) reciprocal space maps (RSM) corresponding to Ge and $Ge_{1-x-y}Si_xSn_y$. These maps exhibit narrow and symmetrical profiles consistent with superior crystallinity of the layers. Note that the $Ge_{1-x-y}Si_xSn_y$ (224) peak falls below the relaxation line (indicated by the red line

$_{x-y}Si_xSn_y$ as the Sn content increases with decreasing temperature. We observe that under these temperature conditions the incorporation of Si in the crystal decreases by half. This is likely due to the reduced reactivity of $Si_4H_{10}$ relative to $SnH_4$ and $GeH_3Cl$ at lower temperatures, limiting its efficiency as a source of Si.

Finally, we note that XPS analysis as a function of layer depth was also conducted on selected samples to further corroborate the Si incorporation and determine the composition. Depth profiling was performed by collecting spectra after subjecting the film to a 25-second sputtering step between each collection. The red and green plots in Figure 2 correspond to representative spectra from sample 1 in the table, illustrating the Sn peaks in the main panel and Si peaks in the inset. The Sn signals are smooth curves, while the Si are slightly noisy but still well-defined, unambiguously illustrating the presence of Si throughout the layer. The integrated peak intensities from the red plots in figure), and it is vertically aligned with the Ge peak along the pseudomorphic direction, indicating that the epilayer is biaxially strained to the buffer and tetragonally distorted from cubic symmetry.

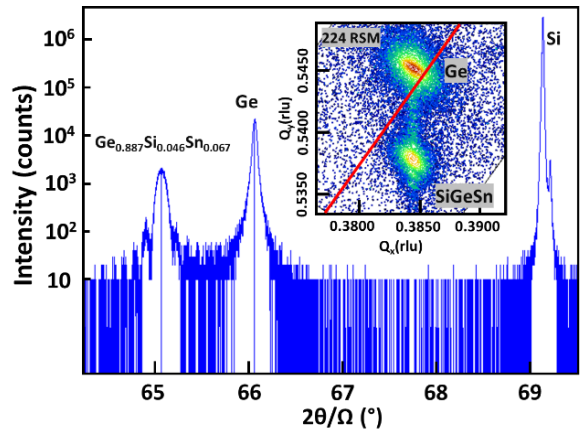

**Figure 3** XRD plots showing 004 peaks (blue line) and (224) RSM (inset) for ~ 6.7% Sn, 4.6% Si sample grown on a Ge buffer. The relaxation line passes slightly below the buffer peak indicating a slight tensile strain. The $Ge_{1-x-y}Si_xSn_y$ (224) peak lies below the relaxation line and aligns with the Ge peak in the vertical direction, indicating pseudomorphic growth.

The horizontal and vertical lattice parameters of the distorted lattice were measured to be $a$ =5.667 Å and $c$= 5.728 Å, respectively. These values are used to calculate the cubic parameter $a_0$ = 5.7024 Å of the sample. Using Vegard's Law and adopting the RBS Si concentration of 4.6%, the above data yield a compressive strain of -0.6051 % and Sn content 6.7% for this sample. Note that the common in-plane parameter of Ge and $Ge_{1-x-y}Si_xSn_y$, $a$ =5.667 Å, is larger than the bulk Ge value $a$ ~ 5.657 Å. This discrepancy is attribute to the slight tensile strain induced in Ge during growth on the thermally mismatched Si.

XRD analysis of all samples in Table 1 provided their relaxed lattice constants, strain states, and Sn concentrations. The latter were typically larger than the RBS values, supporting the claim that the Sn measured by RBS in these samples is fully substitutional. The table shows a systematic increase of the lattice parameter ($a_0$) and compressive strain ($\varepsilon$) with increasing Sn content, as expected.

All samples in this study are compressively strained. No relaxation was observed in any of the films, regardless of composition and thickness. We note that strained structures of $Ge_{1-y}Sn_y$ binary alloys are routinely produced by MBE on Ge wafers and Ge-buffered Si wafers. The ultra-low temperature and pressure conditions of this technique promote the integration of lattice-coherent layers which have been utilized to study fundamental properties and demonstrate device applications. For instance, MBE-grown strained films of $Ge_{1-y}Sn_y$ have been reported to serve as channel materials in MOSFET due to their higher hole mobility.[28, 29] A key advantage of our strained $Ge_{1-x-y}Si_xSn_y$ samples over $Ge_{1-y}Sn_y$ is the ability to controllably add small amounts of Si whose incorporation allows for fine-tuning of optical and electrical properties as well as strain states, providing additional flexibility in device design.

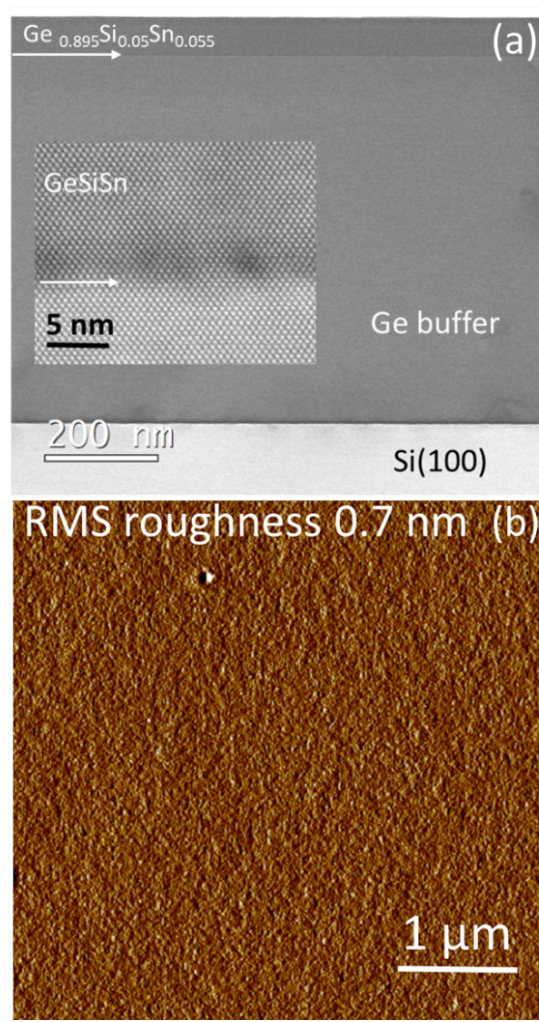

**Figure 4** (a) Bright field XTEM image of 70 nm thick $Ge_{0.895}Si_{0.05}Sn_{0.055}$ film grown on 700 nm thick, Ge buffer. The top layer is flat and crystalline. Inset shows a high-resolution view of the interface between film and buffer. (b) AFM image of the same sample showing a smooth surface with very low RMS roughness of 0.7 nm.

Further structural characterizations were conducted using XTEM. Figure 4 (a) shows a phase contrast image of the 5.5 % Sn and 5% Si alloy sample 1 in Table 1 acquired in scanning mode (XSTEM) with a JEOL ARM200F microscope. The image highlights the excellent crystallinity and morphology observed in these samples. The $Ge_{1-x-y}Si_xSn_y$ top layer is uniform, monocrystalline, and defect-free as expected due to the seamless integration of the film on Ge facilitated by the pseudomorphic growth. The inset shows a high-resolution image revealing a sharp, epitaxial interface between the film and Ge buffer. The surface profile of the film is planar within the field of view. AFM examinations corroborated the flat morphology as shown Figure 4(b) which displays an AFM image of the same sample. The surface is smooth throughout and lacks cross hatch patterns typical for relaxed Ge$_{1-y}$Sn$_y$ layers grown on Ge buffers. The AFM RMS roughness is 0.7 nm, which is typical for these fully strained samples. Additionally, Nomarski optical images confirmed the flat surface morphology across larger area layouts.

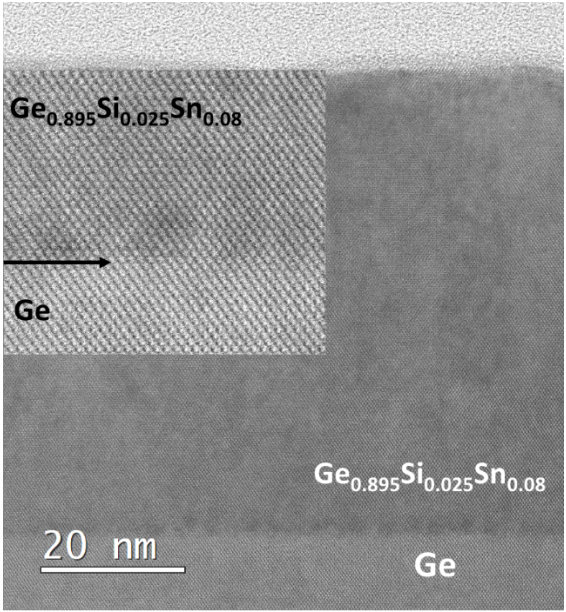

**Figure 5** XTEM bright field image of 60 nm thick Ge$_{0.895}$Si$_{0.025}$Sn$_{0.08}$ film grown on 600 nm thick Ge buffer. Inset shows a high-resolution image of the interface between epilayer and Ge buffer.

Extensive TEM characterizations were conducted on a set of four samples with varying stoichiometries as listed in the table to fully assess the crystallinity of these materials. Figure 5 shows images of a film with Ge$_{0.895}$Si$_{0.025}$Sn$_{0.08}$ composition and a thickness of 55 nm grown upon a 600 nm Ge buffer. The main panel image displays the entire thickness profile of the film, highlighting the dearth of defects as expected due to the pseudomorphic growth. The inset presents a high-resolution image of the interface between Ge and the strained epilayer revealing the defect-free epitaxial relationship and the full commensuration of the (111) lattice fringes. The TEM-measured layer thickness is 60 nm, close to the 55 nm obtained by RBS. In all cases the microstructure consistently resembled the one shown in Figure 5, irrespective of Sn concentration and film thickness.

**Growth of Ge$_{1-y}$Sn$_y$ alloys on Ge buffers and bare Si wafers via GeH$_3$Cl**

Our previous work with GeH$_3$Cl demonstrated its potential as a CVD precursor for fabricating Ge epitaxial layers on Si and photodiode devices.[27] We also conducted proof of concept experiments aimed at synthesizing Ge$_{1-y}$Sn$_y$ via reactions of GeClH$_3$ with SnD$_4$ at ultra-low temperatures (T<200 °C) using the GSME method. Initial results indicated that dilute amounts of Sn (up to 2%) were incorporated into Ge, as evidenced by XRD.[27] In this present study we expanded our efforts to produce a set of Ge$_{1-y}$Sn$_y$ films with a wide range of alloy compositions, and to characterize their materials properties in detail. The synthesis of Ge$_{1-y}$Sn$_y$ was conducted on both Ge buffers and on pristine Si wafers through reactions of GeClH$_3$ and SnH$_4$, following the same methods used for the above Ge$_{1-x-y}$Si$_x$Sn$_y$ on Ge samples. An objective was to test the limits of GeH$_3$Cl for the synthesis of Ge$_{1-y}$Sn$_y$ binaries under experimental conditions similar to the ones employed above for the ternaries and compare results between the two systems.

For each run, the substrates were chemically cleaned and placed in the GSME chamber, where they were degassed on the wafer stage prior to growth. Pure GeH$_3$Cl vapor was introduced first into the chamber followed by gaseous SnH$_4$ without a carrier gas. The reactant pressure was adjusted between 4.5 x10$^{-5}$ and 7x10$^{-5}$ Torr and the temperature varied from 200 °C to 175 °C, allowing for an increase in Sn contents from 2.1 to 7.5%. Under these conditions, we produced films with smooth mirror-like surfaces on both Ge buffers and on Si wafers. Extensive characterizations were performed to assess the material properties. Table 2 lists the growth temperatures, compositions, lattice constants, strain states and layer thicknesses for selected samples.

| Sample | Tg (°C) | XRD Sn (%) | RBS Sn (%) | ε (%) | $a_0$ (Å) | Thickness (nm) |
|---|---|---|---|---|---|---|
| 1 | 230 | 2.1 | 2.2 | -0.1382 | 5.674 | 66 |
| 2 | 210 | 3.2 | 3.2 | -0.3338 | 5.684 | 52 |
| 3 | 195 | 3.9 | 3.9 | -0.4379 | 5.6892 | 29 |
| 4 | 175 | 7.5 | 7.0 | -0.9461 | 5.7198 | 28 |

**TABLE 2.** Representative $Ge_{1-y}Sn_y$ (*y*=0.022-0.70) samples grown on Ge-buffers between 230 and 175 °C. Sample properties, including film thicknesses, Sn concentrations (determined from RBS and XRD) relaxed lattice parameters ($a_0$) and strain states (ε) obtained from the XRD, are listed.

The Sn content and thickness of the $Ge_{1-y}Sn_y$ films were determined by RBS. Ion channeling was employed to examine the epitaxial alignment and evaluate bulk crystallinity. Backscattering intensity profiles of the aligned spectra were obtained, resembling those in Figure 1(b) above, indicating the presence of monocrystalline layers with low-defectivity and fully substitutional Sn contents. HRXRD measurements corroborated the RBS compositions and revealed that all films are pseudomorphic and compressively strained. Figure 6 presents the XRD plots for a 3.2 % Sn film described in the second row of the table. The on-axis plot (blue line) shows sharp (004) peaks for $Ge_{0.968}Sn_{0.032}$ and Ge, indicating good crystallinity and epitaxial alignment along the growth direction. The corresponding 224 RSM plots are included in the inset. The peaks align well along the vertical direction indicating that $Ge_{0.968}Sn_{0.032}$ is lattice matched to Ge within the plane of growth and fully strained to the buffer layer. The in-plane (*a*) and vertical (*c*) lattice parameters for all films in this study were determined from RSM plots and used to calculate the cubic parameters ($a_0$) listed in Table 2. Sn contents were then obtained using Vegard's law and the values correlate well with RBS results, showing good agreement between the two techniques.

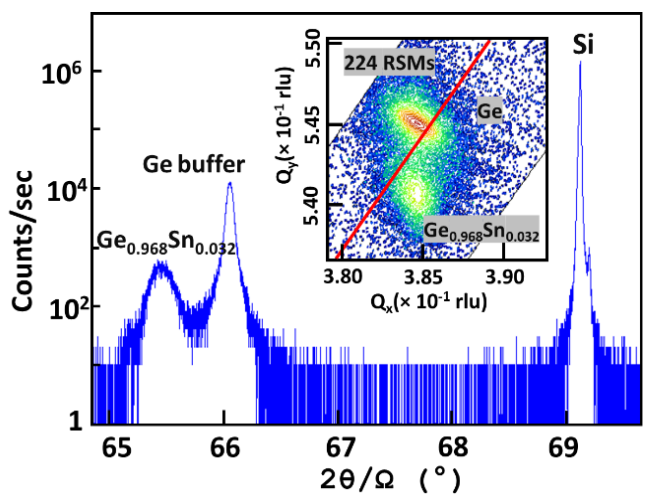

**Figure 6** XRD spectra of $Ge_{0.968}Sn_{0.032}$ grown on Ge. Main panel displays on-axis plot (blue trace) showing (004) peaks. Inset shows (224) RSM of same material. The relaxation line (red) in RSM, passes below Ge peak indicating a residual tensile strain in the buffer layer. The $Ge_{0.968}Sn_{0.032}$ peak aligns with the Ge peak in the vertical direction indicating identical in-plane lattice parameters.

XTEM images for the 2.1 % Sn film described in the first row of the table are shown in Figure 7. Panel (a) illustrates the full epilayer profile, showing no sign of defects, as expected due to the pseudomorphic nature of the crystal. Additionally, no evidence of Sn segregation is observed, corroborating the single-phase character of the alloy. Panel (b) presents a high-resolution image of the interface region between the epilayer and buffer. The in-plane lattice-matching in this case ensures a sharp and uniform heterojunction that is devoid of

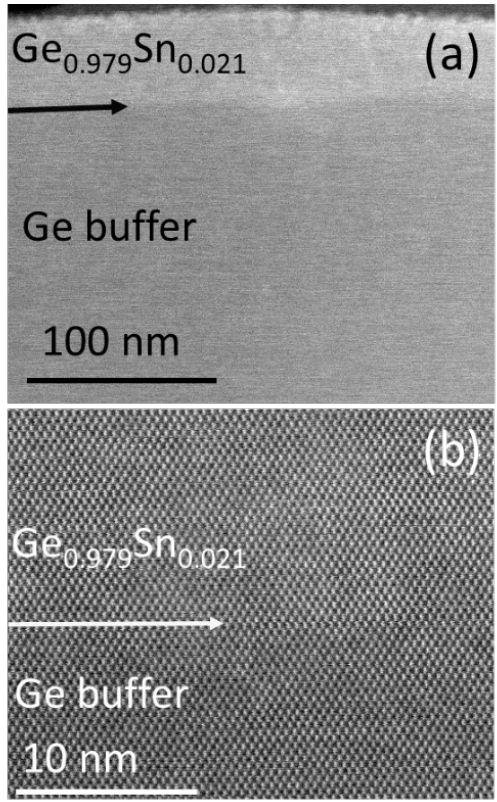

**Figure 7** (a) XTEM images of a 66 nm thick Ge$_{0.979}$Sn$_{0.021}$ layer on a Ge buffer. (b) High resolution image of the interface between the two materials.

defects and exhibits fully commensurate (111) lattice fringes. The TEM images also show that the free surface is slightly etched. This etching occurred during thinning the layer with focused ions beams to fabricate electron transparent specimens for TEM experiments, and does not represent the original surface of the layer.

The above results show that Ge$_{1-y}$Sn$_y$ films were grown fully strained to Ge, but with reduced Sn compositions ranging from ~ 2 - 7 % and significantly lower thicknesses relative to Ge$_{1-x-y}$Si$_x$Sn$_y$ (see table 2). This observation promoted further exploration of the growth behavior of Ge$_{1-y}$Sn$_y$ using the GeH$_3$Cl method, bypassing the Ge buffer and producing samples directly on Si. The motivation behind this approach is that growth on Si promotes relaxation of the misfit strain thereby enabling a wider range of Sn incorporation and larger film thickness approaching bulk levels. The growth experiments were conducted following the same protocols and methods described preciously for Ge$_{1-y}$Sn$_y$ and Ge$_{1-x-y}$Si$_x$Sn$_y$ films on Ge buffers.

A series of Ge$_{1-y}$Sn$_y$ samples with mirror like surfaces were produced directly on Si and their material properties were determined. A summary of representative films is listed in Table 3 along with relevant characterizations results. The data show that the Sn concentrations increase from 2.4% to 7.2% as the temperature decreases from 200°C to 170°C, in analogy to layers grown on Ge buffers. A notable distinction is that the XRD-derived Sn contents are slightly lower than those measured by RBS. The discrepancy is largest for the first two samples and diminishes with decreasing temperature, as is apparent from the closely matched XRD/RBS Sn contents of the fourth sample. While the Ge$_{1-y}$Sn$_y$ samples on Ge are fully strained, the samples on Si are mostly relaxed, as expected, exhibiting residual compressive strains that depend on composition and growth temperature.

The crystallinity of the samples on Si was slightly inferior compared to those on Ge. This is due the mismatch-induced interface defects, whose dislocation cores propagate through the layer, degrading the microstructure. Nevertheless, the films were monocrystalline and epitaxial, indicating that this approach may hold value in technologies requiring direct growth on Si, under very low thermal budgets, using straightforward deposition techniques. Efforts to increase concentration and thickness of these samples resulted in non-homogeneous multi-phase materials containing large amounts of interstitial Sn impurities, indicating that only a limited range of $Ge_{1-y}Sn_y$ compositions can be achieved using the $GeH_3Cl$ approach at ultra-low temperatures. The compositions and thicknesses of $Ge_{1-y}Sn_y$ on Ge and on Si are virtually identical within the margin of error.

| Sample | Tg (°C) | XRD Sn (%) | RBS Sn (%) | ε (%) | a$_0$ (Å) | Thickness (nm) |
|---|---|---|---|---|---|---|
| 1 | 200 | 2.5 | 3 | -0.1706 | 5.6778 | 48 |
| 2 | 197 | 3.1 | 3.6 | -0.1208 | 5.6829 | 69 |
| 3 | 192 | 3.9 | 4.3 | -0.0471 | 5.6899 | 35 |
| 4 | 173 | 7.2 | 7.3 | -0.4546 | 5.7188 | 27 |

**TABLE 3.** Ge$_{1-y}$Sn$_y$ (y=0.03-0.73) samples grown directly on Si between 200 and 173 °C. Growth temperatures (Tg), RBS/XRD Sn concentrations, relaxed lattice parameters (a$_0$), residual compressive strains (ε), and film thicknesses are listed.

**Raman studies**

Raman spectra were obtained for a series of $Ge_{1-x-y}Si_xSn_y$ films described above to further investigate bonding properties of these materials and compare the results to those obtained from other methods. For the $Ge_{1-y}Sn_y$ binaries, our various characterizations clearly showed that the samples were single-phase alloys. However, for the ternaries, the atomic distribution in the average diamond lattice represents an additional variable that may depend on the particular method of synthesis and on the chemical precursors used for the growth. Therefore, our Raman investigation here focuses specifically on these ternary samples since the materials were grown using $GeH_3Cl$ for the first time.

The Raman spectra were collected in the backscattering configuration in a custom micro-Raman setup using 1.3 mW of 532 nm illumination. The laser light was focused on the sample using a 100× objective and the scattered light collected with a standard CCD detector. The Raman shifts in the alloys were carefully calibrated by also measuring bulk Ge and Si samples and rescaling the results so that the peaks agree with high accuracy measurements from Trzeciakowski and coworkers,[30] which imply $\omega_{Si}$ = 520.69 cm$^{-1}$ and $\omega_{Ge}$ = 300.265 cm$^{-1}$ at room temperature. Spikes in the collected spectra were removed using a custom variant of the Katsumoto-Ozaki algorithm,[31] in which we combine binomial smoothing with Savitzky-Golay smoothing.

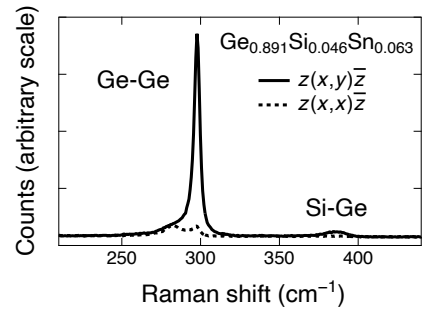

**Figure 8** Raman spectrum of a GeSiSn alloy obtained with 532 nm excitation. The polarization conditions are indicated in the Porto notation, where $x$, $y$, $z$ are the directions of the cubic crystalline axes. The solid line corresponds to allowed first-order Raman scattering and the dashed line to a forbidden configuration. The low-energy peak at 282 cm$^{-1}$, quite apparent in the "forbidden" spectrum, is a two-phonon feature that is also observed in pure Ge.

Figure 8 shows a typical Raman spectrum from one of the ternary samples. For the chosen backscattering configuration, the spectrum consists of longitudinal optical vibrations. The superb crystalline quality is demonstrated by the excellent depolarization ratio, in full accordance with the selection rules for diamond-structure materials. The Raman spectrum from ternary $Ge_{1-x-y}Si_xSn_y$ alloys was first studied by D'Costa and coworkers.[32] For relatively modest Si and Sn concentrations it is quite similar to the Raman spectrum of Ge-rich SiGe alloys, dominated by two peaks that are informally referred to as "Ge-Ge" and "Si-Ge" vibrations. "Si-Sn" vibrations may also contribute to the Raman spectrum, but they overlap in frequency with the Si-Ge modes, since Si is much lighter than Ge or Sn and the Raman frequency is approximately proportional to $\mu^{-1/2}$, where $\mu$ is the reduced mass of the two atoms.[33] The intensities of these peaks are roughly proportional to the fraction of Ge-Ge and Si-Ge bonds in the sample. We see clear evidence for the Ge-Ge and Si-Ge modes in Figure 8, but the latter is much weaker on account of the low Si-concentrations. This also explains why no Si-Si Raman peak is observed.

Alloy Raman peaks have a characteristic asymmetric profile, and the frequency of a mode is phenomenologically defined as the maximum of its Raman peak. For the Si-Ge mode, this frequency has a complicated compositional dependence, but the Ge-Ge mode frequency is a linear function of $x$ and $y$ that lends itself nicely to characterization work, including composition determinations. Accordingly, we focus our analysis on this peak, which is shown in more detail and compared with the Raman peak of pure Ge in Figure 9. To account for the asymmetric profile, we fit our Raman peaks with an exponentially modified Gaussian (EMG) lineshape,[32] from which we extract the peak maxima.

D'Costa and coworkers[32] proposed a compositional dependence of the LO mode frequency given by

$$\omega_{Ge-Ge}(x,y) = \omega_0^{Ge} - \alpha_{Ge-Ge}^{GeSi}x - \alpha_{Ge-Ge}^{GeSn}y + b_{Ge-Ge}\epsilon(x,y) \tag{3}$$

In the last term in Eq. (3), $\epsilon(x,y)$, is the strain that appears in Table I. For a tetragonal distortion, the strain coefficient is given by $b_{Ge-Ge} = \frac{1}{\omega_{Ge-Ge}}\left[q - p\left(\frac{C_{12}}{C_{11}}\right)\right]$, where $C_{11}$ and $C_{12}$ are elastic constants, and we use the notation $p, q, r$ for the three symmetry-allowed anharmonic coefficients in the diamond structure.[34] Using this notation, the Grüneisen parameter for the mode, defined as $\gamma = -\partial \ln \omega_{Ge-Ge}/\partial \ln V$, is given by $\gamma = -(p+2q)/6\,\omega_{Ge-Ge}^2$. D'Costa et al used $b_{Ge-Ge} = -415$ cm$^{-1}$ based on $p$ and $q$ values for pure Ge from a classic paper by Cerdeira et al.[35] These values, however, lead to $\gamma = 0.88$, considerable less than the experimental value $\gamma = 1.00$.(Ref. 36) In more recent direct Raman measurements on epitaxially strained Ge$_{1-x}$Si$_x$ epitaxial films, Reparaz et al. found $b_{Ge-Ge} = -(460\pm20)$ cm$^{-1}$ for pure Ge, in much better agreement with the experimental Grüneisen parameter.[37] They also found a very weak dependence on composition for $b_{Ge-Ge}$, justifying the use of a single constant in Eq. (3). Very similar results ($b_{Ge-Ge} = -450\pm30$ cm$^{-1}$ for pure Ge) were obtained by Pezzoli and co-workers, who also discuss a theoretical model supporting the very weak compositional dependence.[38] However, in more recent work Yokogawa et al.(Ref. 39) claim a stronger strain dependence ($b_{Ge-Ge} \sim -575$ cm$^{-1}$). In view of these uncertainties, we decided to carry out our own measurements on pure Ge epitaxial layers grown in our lab, and then use the value of $b_{Ge-Ge}$ determined from those measurements to analyze our GeSiSn data. We reasoned that even if Raman strain measurements contain some unknown systematic error that explains the variations from group to group, this error is likely to cancel out if we use our own strain measurements in pure epitaxial Ge to correct our GeSiSn results for strain, since both sets of data are obtained using the same experimental setup under identical conditions. The value we found was $b_{Ge-Ge} = -427\pm30$ cm$^{-1}$. This is much more consistent with the results from Reparaz and Pezzoli,[37, 38] than with Yokogawa.[39]

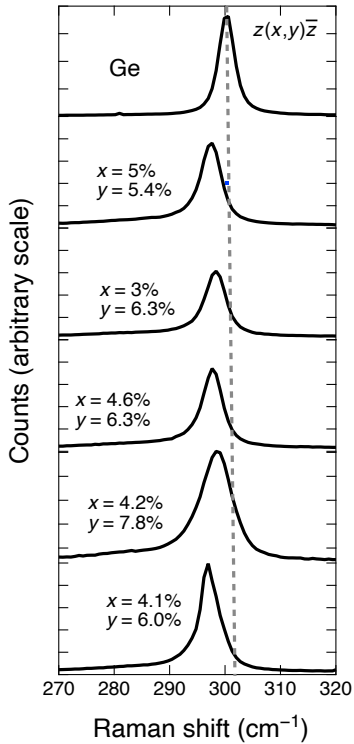

**Figure 9** Allowed $z(x,y)\bar{z}$ Raman spectra from selected chlorogermane GeSiSn samples around the frequency corresponding to the Ge-Ge mode.

D'Costa and co-workers did a two-dimensional adjustment of Eq. (3) to their experimental data, so that the two linear coefficients $\alpha_{Ge-Ge}^{GeSi}$ and $\alpha_{Ge-Ge}^{GeSn}$ could be extracted from the fit. Their main purpose was to verify the conjecture that these coefficients should be "transferable" from the binary alloys $Ge_{1-x}Si_x$ and $Ge_{1-y}Sn_y$ to the ternary $Ge_{1-x-y}Si_xSn_y$. The conjecture turned out to be correct within experimental error. However, since $\alpha_{Ge-Ge}^{GeSi}$ is about four times smaller than $\alpha_{Ge-Ge}^{GeSn}$, Si concentrations approaching at least 20% are needed for the fits to converge well. These are much higher than the Si concentrations in the present paper. Furthermore, our main interest here is to assess the structural properties of $Ge_{1-x-y}Si_xSn_y$ alloys synthesized with GeH$_3$Cl compared to ternaries grown using the more traditional polygermane route (Ge$_4$H$_{10}$, Ge$_3$H$_8$). For this purpose, we simply subtract the strain shift and the Si-induced shift using the value $\alpha_{Ge-Ge}^{GeSi}$= 17.1 cm$^{-1}$ from D'Costa *et al*, and consider only the Sn-concentration dependence of the Raman frequencies

$$\Delta\omega_{Ge-Ge}(y) = \omega_{Ge-Ge}(x,y) + \alpha_{Ge-Ge}^{GeSi}x - b_{Ge-Ge}\epsilon(x,y) - \omega_0^{Ge}.$$

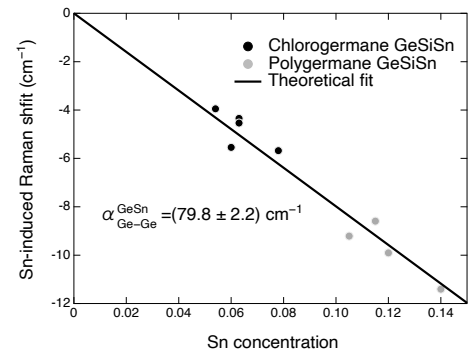

**Figure 10** Sn-dependence $\Delta\omega_{Ge-Ge}(y)$ of the Ge-Ge mode frequency in GeSiSn alloys. The solid line is a linear fit. in pure Ge.

Figure 10 shows $\Delta\omega_{Ge-Ge}(y)$ for selected ternary alloys grown with GeH$_3$Cl and for other ternary alloys grown with polygermanes. We see that the data is consistent with a linear *y*-dependence and that there seems to be no distinction between the two synthetic approaches to GeSiSn alloys, corroborating the conclusion that the GeH$_3$Cl route leads to completely equivalent materials. A fit of the *y*-dependence with the expression $-\alpha_{Ge-Ge}^{GeSn}y$ gives $\alpha_{Ge-Ge}^{GeSn} = (79.8\pm2.2)$ cm$^{-1}$, somewhat below the published values $\alpha_{Ge-Ge}^{GeSn} = (94.\pm2.2)$ cm$^{-1}$ (Ref. 29), $\alpha_{Ge-Ge}^{GeSn} = (94.0\pm7.1)$ cm$^{-1}$ (Ref. 40), and $\alpha_{Ge-Ge}^{GeSn} = (89.9\pm2.9)$ cm$^{-1}$ (Ref. 41), but in even better agreement with the transferability conjecture in Ref. 32, since the corresponding measurement in binary $Ge_{1-y}Sn_y$ alloys gives $\alpha_{Ge-Ge}^{GeSn} = (75.4\pm4.5)$ cm$^{-1}$ (Ref. 42).

**SUMMARY:**

In this paper we demonstrate the potential of a novel GeH$_3$Cl precursor to replace higher order germanes such as Ge$_4$H$_{10}$ and Ge$_3$H$_8$ for low-temperature, CMOS-compatible development of Sn/Si -based semiconductors. To highlight its effectiveness, we used CVD to fabricate $Ge_{1-x-y}Si_xSn_y$ epitaxial films on Ge buffered Si substrates. This process involved interactions of GeH$_3$Cl with SnH$_4$ and Si$_4$H$_{10}$ between 200°C and 160°C. These ultra-low temperature conditions are typically associated with MBE processes using solid-source fluxes. Achieving synthesis of monocrystalline films under these conditions via a purely chemical process based on molecular reactions is unique and represents an advance in crystal growth of group IV alloys.

The $Ge_{1-x-y}Si_xSn_y$ films were fully strained to Ge buffers, exhibiting Si contents of 3-5% Si and Sn contents of 5-11 % Sn with thicknesses up to 100 nm. For comparison, $Ge_{1-y}Sn_y$ binaries were also grown on Ge buffers under the same conditions. These films are also fully strained to Ge, however, they contain significantly lower amounts of Sn and have reduced thicknesses relative to the ternaries. We next attempted the growth of $Ge_{1-y}Sn_y$ directly on Si without buffers to mitigate the misfit strain with the intention to promote formation of thicker layers with high Sn contents. The $Ge_{1-y}Sn_y$ films on Si exhibited mostly relaxed strain states, as expected, but their thicknesses and Si contents were low and comparable to those on Ge.  Attempts to increase Sn content and thickness beyond a limited range were unsuccessful resulting in multiphase products.   The study demonstrates that ternary alloys produced via the GeH$_3$Cl method can incorporate much higher Sn contents than binary analogs regardless of strain.  This suggests that the small Si atoms might compensate for the large Ge-Sn mismatch by lowering bond strains in Ge$_{1-x-y}$Si$_x$Sn$_y$ compared to Ge$_{1-y}$Sn$_y$.  This, in turn, promotes larger thicknesses and induces better crystallinity.

## ACKNOWLEDGMENT

This work was supported by the U.S. National Science Foundation under Grant Nos. DMR-2119583 and DMR-2235447 and by the Air Force Office of Scientific Research under Grant No. FA9550-23-1-0285.

## CONFLICT OF INTEREST

The authors have no conflicts to disclose

## DATA AVAILABILITY

The data that supports the findings of this study are available within the article. Further questions about supporting data should be send to the corresponding author.

## CREDIT STATEMENT

This article has been submitted to the Journal of Vacuum Science and Technology A. After it is published, it will be found at [Link](Link).